\documentclass[prb,twocolumn,nopacs,amsmath,amssymb]{revtex4}

\newcommand{\bk}{{\bf k}}

\newcommand{\bq}{{\bf q}}

\newcommand{\br}{{\bf r}}
\newcommand{\brp}{{{\bf r}^{\prime}}}

\newcommand{\ef}{\epsilon_{\rm F}}
\newcommand{\vf}{v_{\rm F}}
\newcommand{\kf}{k_{\rm F}}

\newcommand{\bwt}{\begin{widetext}}
\newcommand{\ewt}{\end{widetext}}

\usepackage{graphics,graphicx,amsmath}

\usepackage{bm}
\usepackage{color}

\begin{document}

\bibliographystyle{apsrev}

\date{\today}

 \author{R. Rold\'{a}n$^1$, J.-N. Fuchs$^{2,3}$ and M. O. Goerbig$^3$}
\affiliation{\centerline{$^1$Instituto de Ciencia de Materiales de
    Madrid,
CSIC, Cantoblanco E28049 Madrid, Spain}\\
\centerline{ $^2$Laboratoire de Physique Th\'eorique de la Mati\`ere Condens\'ee, CNRS UMR 7600, UPMC, 4 place Jussieu, F-75252 Paris}\\
\centerline{$^3$Laboratoire de Physique des Solides, Univ. Paris-Sud, CNRS, UMR 8502, F-91405 Orsay Cedex, France}}

\title{ Collisionless Hydrodynamics of Doped Graphene in a Magnetic Field}

\begin{abstract}
The electrodynamics of a two-dimensional gas of massless fermions in graphene is studied by a  collisionless hydrodynamic approach. A low-energy dispersion relation for the collective modes (plasmons) is derived both in the absence and in the presence of a perpendicular magnetic field. The results for graphene are compared to those for a standard two-dimensional gas of massive electrons. We further compare the results within the classical hydrodynamic approach to the full quantum mechanical calculation in the random phase approximation. The low-energy dispersion relation is shown to be a good approximation at small wave vectors. The limitations of this approach at higher order is also discussed. 
\end{abstract}


\maketitle

\section{Introduction}

Plasmons are collective excitations of the electron liquid that completely dominate its excitation spectrum at long wavelengths.\cite{S67} Although the random phase approximation (RPA) is the simplest theory that can account for them within a quantum mechanical description, plasmons have a classical origin and can be described  qualitatively, at lowest order, by a proper hydrodynamical approach.\cite{F73} The screening properties of standard two-dimensional electron gases (2DEG) of massive electrons, with a parabolic dispersion relation, as well as their collective modes, have been extensively studied.\cite{AFS82,GV05} 

Recently, much attention has been payed to understanding the peculiarities of graphene plasmons.\cite{GPN12} Although graphene also supports plasmon modes with a low-energy $\omega \sim \sqrt{q}$ dispersion relation, where $\omega$ is the energy and $q$ the wave vector, the dependence of the dispersion on the electron density $n$ is different: whereas $\omega\sim \sqrt{nq}$ for a 2DEG, the characteristic linear dispersion relation of graphene leads to a $\omega\sim\sqrt{n^{1/2}q}$ behavior.\cite{S86,WSSG06,HS07} The particle-hole excitation spectrum, defined as the region of the $\omega-q$ space where  electron-hole excitations are possible, is also rather different in the two cases.\cite{RGF10} When the additional effect of an external magnetic field perpendicular to the layer is considered, new differences between the 2DEG and graphene appear, due to the essentially different Landau level quantization of the spectrum in the two cases.\cite{IJFB07,S07,BM08,RFG09} 

In this  paper, we  present a classical hydrodynamic approximation\cite{F73} to study the collective excitations in graphene, including the effect of a finite magnetic field in the spectrum. This classical approach gives a simple description of the origin and dispersion of plasmons, which can be identified with longitudinal density oscillations or sound waves in usual gases and liquids. We obtain an approximate dispersion relation for the plasmon and for the upper hybrid mode, which is the name of the plasmon dressed by the contribution of the magnetic field. We further compare the hydrodynamic results to the full quantum mechanical RPA approximation, and show how the former give a reasonable approximation in the long wavelength limit. Finally, we discuss the limitations of the classical hydrodynamic approximation, and compare our results for relativistic fermions in graphene to the well-studied case of massive electrons in a 2DEG.  

We note that hydrodynamics has already been applied to graphene either in the quantum critical regime of high temperature and low doping of a clean system\cite{MS08} or to describe transport in a disordered and doped system.\cite{BM09,MHS13}  Very recently, the renormalization due to electron-electron interactions of the classical plasmon mode in graphene as well as of the upper hybrid mode has been investigated, within the framework of Landau's Fermi-liquid theory.\cite{LSF13}  As a consequence of the lack of Kohn's theorem in graphene,\cite{RFG10} electron-electron interactions renormalize the cyclotron frequency.\cite{IJFB07,BM08,S10,RFG10}

\section{Hydrodynamical theory of linear response}\label{Sec:Hydro}

Hydrodynamical theory describes electronic motion in terms of two dynamical variables, namely the electron density, $n(\br,t)$ and the electron velocity ${\bf v}(\br,t)=(v_x,v_y,0)$. At zero temperature $T=0$ (in the absence of heat current) the charge and momentum currents are proportional. Euler's and the continuity equations read\cite{F73}

\bwt
\begin{eqnarray}
 \alpha\frac{d}{dt}\left(\frac{ \epsilon_F}{\vf^2} {\bf J}(\br,t)\right)&=&e{\pmb \nabla}P(\br,t)+en(\br,t){\pmb \nabla}\int d\brp \frac{e^2}{|\br-\brp|}[n(\brp,t)-n_0]-e{\bf J}(\br,t)\times{\bf B},\label{Eq:Euler}\\
\frac{\partial n(\br,t)}{\partial t}&=&\frac{1}{e}{\pmb \nabla}\cdot[{\bf J}(\br,t)],\label{Eq:Cont}
\end{eqnarray}
\ewt
where $\epsilon_F$ is the Fermi energy, $e$ is the electron charge, ${\bf J}(\br,t)$ is the charge current (see also Appendix \ref{AppendCons}), $P(\br,t)$ is the pressure in the layer and ${\bf B}=(0,0,B)$ is the magnetic field perpendicular to the plane. Furthermore, the coefficient $\alpha$ is related to the dispersion relation, $\epsilon \propto k^{\alpha}$, which we consider here. This exponent affects the density of states 
\begin{equation}\label{Eq:DOS}
\rho(\epsilon)\propto \epsilon^{d/\alpha-1}, 
\end{equation}
where $d$ is the spatial dimension of the electron system, and it allows us to discuss the 2DEG with a parabolic band dispersion ($\alpha=2$, $\epsilon=k^2/2m$) on an equal footing with graphene ($\alpha=1$, $\epsilon=v_F k$). We take $\hbar\equiv 1$ from now on. Indeed, the quantity intervening on
the left hand side of (\ref{Eq:Euler}) is thus the cyclotron mass,
\begin{equation}\label{Eq:Cmass}
m_c=\alpha\frac{\ef}{\vf^2}.
\end{equation}
We have defined the Fermi velocity as $\vf=\frac{\partial \epsilon}{\partial k}|_F$, in terms of the (massless or massive) electronic dispersion relation $\epsilon(k)$.
The second term on the right hand side of Eq. (\ref{Eq:Euler})  represents the long-range Coulomb interaction among the carriers, and the third term is the Lorentz force term, which accounts for the presence of an external magnetic field. The electron density can be decomposed as
\begin{equation}
n(\br,t)\equiv n_0+\delta n(\br,t)
\end{equation}
where $n_0$ is the mean  average density of the system (which is assumed to be neutralized by a rigid uniform background), and $\delta n(\br,t)$ is the electron-density fluctuation. Rigorously, these equations must be combined with Maxwell's equations. However we neglect retardation effects as the speed of light $c$ is much larger than the Fermi velocity $\vf\approx c/300$. The charge current is defined, in terms of the electron density and velocity, as
\begin{equation}
{\bf J}(\br,t)=-en(\br,t){\bf v}(\br,t).
\end{equation}

We may further simplify the problem by considering the linear response of the initially stationary system to an applied perturbation. Notice that within this approximation, the velocity ${\bf v}$, and the electric (${\bf E}$) and magnetic (${\bf B}$) fields are of first order. Therefore we approximate 
\begin{equation}
\dot{\bf v}=\frac{\partial{\bf v}}{\partial t}+({\bf v}\cdot {\pmb \nabla}){\bf v}\simeq \frac{\partial{\bf v}}{\partial t}
\end{equation}
and 
\begin{equation}
\frac{\partial (\ef {\bf J}/\vf^2)}{\partial t}\simeq \frac{\ef}{\vf^2} \frac{\partial  {\bf J}}{\partial t} 
\end{equation}
where we have taken the Fermi energy to be approximately time- and position-independent.
In order to related the pressure to the electronic density, we make use of the equation of states, which can be obtained from 
$P(\br,t)=n(\br,t)(\ef-\langle \epsilon\rangle)$, where $\langle \epsilon\rangle =\int_0^{\ef}d\epsilon\,\epsilon\rho(\epsilon)/\int_0^{\ef}d\epsilon\,\rho(\epsilon)$, in terms of the 
density of states (\ref{Eq:DOS}). This yields the general $T=0$ equation of states
\begin{equation}\label{Eq:P}
P(\br,t)=\frac{1}{1+d/\alpha}n(\br,t)\ef.
\end{equation}
Notice that the Fermi energy $\ef$ is itself a function of density, $\ef\propto n^{\alpha/d}$, as one may see from Eq. (\ref{Eq:DOS}).
Whereas for the 2DEG, this yields the usual relation $P=n\ef/2$, where $\ef=\kf^2/2m=\pi n/m$, one obtains for the electronic quantum pressure in graphene
\begin{equation}
 P(\br,t)=\frac{1}{3}\vf\sqrt{\pi}n^{3/2}(\br,t),
 \end{equation}
in terms of the density alone, where we have used $\ef=\vf \kf=\vf\sqrt{\pi n}$. One thus obtains for the pressure gradient in Eq. (\ref{Eq:Euler})
\begin{eqnarray}
{\pmb\nabla} P(\br,t)&=&\frac{\partial P}{\partial n}{\pmb\nabla}\delta n(\br,t)\nonumber\\
&\simeq&\frac{\alpha}{d}\ef{\pmb \nabla}\delta n(\br,t),
\end{eqnarray}
where the approximation in the second line consists of considering an average Fermi energy that is constant in space and time, that is we consider only first-order terms in $\delta n(\br,t)$.

This allows us to write the linearized equations of motion (\ref{Eq:Euler})-(\ref{Eq:Cont}) as 

\begin{eqnarray}
 \alpha\frac{\ef}{\vf^2}\frac{\partial{\bf J}(\br,t)}{\partial t}&=&\frac{\alpha}{d}e\ef{\pmb \nabla}\delta n(\br,t)+en_0{\pmb \nabla}\int d\brp \frac{e^2}{|\br-\brp|}\delta n(\brp,t)\nonumber\\
&&-e{\bf J}(\br,t)\times{\bf B},\label{Eq:Euler2}\\
\frac{\partial \delta n(\br,t)}{\partial t}&=&\frac{1}{e}{\pmb \nabla}\cdot{\bf J}(\br,t)\label{Eq:Cont2}
\end{eqnarray}

Moreover, the \textit{two-dimensional} electric field is determined by the scalar and vector fields
\begin{equation}
{\bf E}(\br,t)=-{\pmb\nabla} \phi(\br,t)-\frac{1}{c}\frac{\partial {\bf A}}{\partial t}
\end{equation}
where $\bf A$ is the vector potential and $\phi(\br,t)=\phi_{ind}(\br,t)$ is the induced potential caused by the excess or deficit of carriers. Furthermore, the electric potential must satisfy the usual wave equation 
\begin{equation}\label{waveeq}
\left[ \nabla^2-\frac{1}{c^2}\frac{\partial^2}{\partial t^2}\right]\phi(\br,t)=-4\pi \rho(\br,t).
\end{equation}
Then, taking into account that $\rho(\br,t)=\rho_{ind}(\br,t)=-e\delta n(\br,t)$ and neglecting retardation effects we can write the Laplace equation as
\begin{equation}
\nabla^2 \phi(\br,t)=4\pi e \delta n(\br,t)
\end{equation}
which can be expressed, after a Fourier transformation, as
\begin{equation}\label{FTphi}
-q^2\phi(\bq,\omega)=4\pi e\delta n(\bq,\omega).
\end{equation}
Substitution of Eq. (\ref{FTphi}) in Eq. (\ref{Eq:Cont}) yields

\begin{eqnarray}
-i\omega\alpha \frac{\ef}{\vf^2}{\bf J}(\bq,\omega)&=&i\frac{\alpha}{d}e\ef\bq\delta n(\bq,\omega)+ien_0\bq\delta n(\bq,\omega)v^{2D}(\bq)\nonumber\\
&&-e{\bf J}(\bq,\omega)\times {\bf B}\label{Eq:FTEuler}\\
-i\omega\delta n(\bq,\omega)&=&\frac{1}{e}i \bq\cdot {\bf J}(\bq,\omega)\label{Eq:FTCont}
\end{eqnarray}
where 
\begin{equation}
v^{2D}(\bq)=\frac{2\pi e^2}{\epsilon_b q}
\end{equation}
is the 2D Fourier transformation of the three-dimensional Coulomb interaction, and $\epsilon_b$ is the background dielectric constant. Notice that, using Eq. (\ref{FTphi}), we can further write
\begin{equation}
-i\bq \phi(\bq,\omega)=i\frac{\bq}{q^2}4\pi e\delta n(\bq,\omega)={\bf E}(\bq,\omega)
\end{equation}
which allows us to express Eqs. (\ref{Eq:FTEuler})-(\ref{Eq:FTCont}) in the more convenient form:
\begin{eqnarray}
-i\alpha\omega\frac{\ef}{\vf^2}{\bf J}(\bq,\omega)&=&i\frac{\alpha}{d}e\ef\bq\delta n(\bq,\omega)+q\frac{n_0e^2}{2\epsilon_b}{\bf E}(\bq,\omega)\nonumber\\
&&-e{\bf J}(\bq,\omega)\times {\bf B}\label{Eq:FTEuler-2}\\
-i\omega\delta n(\bq,\omega)&=&\frac{1}{e}i \bq\cdot {\bf J}(\bq,\omega)\label{Eq:FTCont-2}
\end{eqnarray}

Without loss of generality we can choose $\bq\equiv (q,0)$, which implies that $E_y(\bq,\omega)=0$. Therefore we have, taking into account that ${\bf J}\times {\bf B}=B(J_y,-J_x)$, that
\begin{eqnarray}
-i\alpha\omega\frac{\ef}{\vf^2} J_x(q,\omega)&=&iq\frac{\alpha}{d}e\ef\delta n(q,\omega)+q\frac{n_0e^2}{2\epsilon_b}E_x(q,\omega)\nonumber\\
&&-eBJ_y(q,\omega)\\
-i\alpha\omega\frac{\ef}{\vf^2} J_y(q,\omega)&=&+eBJ_x(q,\omega)\\
-\omega\delta n(q,\omega)&=&\frac{1}{e}qJ_x(q,\omega).
\end{eqnarray}

The above set of equations allows us to write
\begin{equation}
J_x(q,\omega)\left [\omega^2-\frac{1}{d}\vf^2q^2-\omega_c(\ef)^2\right]=i\omega q\frac{n_0e^2\vf^2}{2\alpha\ef\epsilon_b}E_x(q,\omega)
\end{equation}
where we have introduced the energy (or carrier density) dependent cyclotron frequency
\begin{equation}
\omega_c(\ef)=\frac{eB\vf^2}{\alpha\ef}=\frac{eB}{m_c},
\end{equation}
in terms of the cyclotron mass (\ref{Eq:Cmass}). By using the relations between $\bf J$ and $\bf E$ through the conductivity tensor $\pmb\sigma$, $J_x=\sigma_{xx}E_x$ and $J_y=\sigma_{xy}E_x$, then we can finally write the longitudinal and transverse (Hall) conductivities as
\begin{eqnarray}
\sigma_{xx}&=&\frac{i\frac{n_0e^2}{2\epsilon_b m_c}\omega q}{\omega^2-\vf^2q^2/d-\omega_c(\ef)^2}\\
\sigma_{xy}&=&\frac{\frac{n_0e^2}{2\epsilon_b m_c}\omega_c(\ef)q}{\omega^2-\vf^2q^2/d-\omega_c(\ef)^2}
\end{eqnarray}
Notice that in the $B\rightarrow 0$ limit the transverse conductivity $\sigma_{xy}$ vanishes, as it should. 

We now study the upper-hybrid (UH) mode, which is the classical 2D plasmon collective excitation {\it dressed} by the contribution due to the external magnetic field. Its dispersion relation is found by looking for the zeroes of the dielectric function
\begin{equation}
\epsilon_{xx}=1+\frac{4\pi i}{\epsilon_b \omega}\sigma_{xx}
\end{equation}
from which we obtain our final result
\begin{eqnarray}\label{Eq:UH}
\omega_{\rm uh}(q)&=&\sqrt{\omega_c^2+\omega_{p,cl}^2+\omega_s^2}\nonumber \\
&=& \sqrt{\left(\frac{eB}{m_c}\right)^2+\frac{2\pi e^2n_0}{\epsilon_b m_c}q+v_s^2q^2}.
\end{eqnarray}
In the above equation we have introduced the (first) sound frequency $\omega_s=v_s q$ and velocity $v_s$, which is written in terms of the Fermi velocity as
$v_s=\vf/\sqrt{d}$ and that dominates the quadratic term arising from the quantum pressure of the gas. The sound velocity is thus independent of the particular band dispersion (that is independent of $\alpha$) and depends solely on the dimensionality. The information about the band dispersion is encoded in the first and second term, namely in the cyclotron mass -- whereas it is independent of the density in the case of the 2DEG, it scales as $\sqrt{n_0}$ in graphene, as may be seen from Eq. (\ref{Eq:Cmass}). To distinguish between the $B=0$ and the $B\ne 0$ contributions, we can write Eq. (\ref{Eq:UH}) as
\begin{equation}\label{Eq:Plasmons}
\omega_{uh}^2(q)=\omega_c^2+\omega_p^2(q),
\end{equation}
where the cyclotron frequency makes this mode gapped when an external magnetic field is applied perpendicular to the layer, and the approximate expression for the zero-field plasmon dispersion $\omega_p(q)$ is
\begin{equation}\label{Eq:PlasmonB0}
\omega_p(q)\simeq\sqrt{\omega_{p,cl}^2+\gamma\vf^2 q^2},
\end{equation} 
where 
\begin{equation}
\omega_{p,cl}^2=\frac{2\pi e^2 n_0}{\epsilon_b m_c}q
\end{equation}
is the (square of the) classical plasma frequency. Notice that, both for $\alpha=1$ (graphene) and $\alpha=2$ (2DEG), one finds 
$\omega_{p,cl}^2=2e^2\ef q/\epsilon_b$.\footnote{This is related to having $\kf^2=2\pi n$ in the 2DEG and $\kf^2=\pi n$ in graphene as a result of the twofold valley degeneracy in the latter.} Furthermore, we have written the corrective term generally in terms of a parameter $\gamma$, in Eq. (\ref{Eq:PlasmonB0}) to prepare the discussion below about the limits of the collisionless hydrodynamic approach, which has given $\gamma=1/2$ for both graphene and the 2DEG.

For graphene, the dispersion relation (\ref{Eq:UH}) is also contained in the work by Shizuya.\cite{S07} He derived an effective long wavelength gauge theory for graphene in a perpendicular magnetic field. To see the connection, we provide the following dictionary between his notations and ours: $\omega_{\rm eff}\approx eB\vf^2/\ef$ is our cyclotron frequency $\omega_c$ when the Landau-level filling factor $\nu\gg1$, $\alpha_m/\alpha_e$ is given by $v_s^2=\vf^2/2$ in terms of the Fermi velocity, $\bar{\rho}$ is the equilibrium density $n_0$ and $l$ is the magnetic length $l_B=1/\sqrt{eB}$. Then the pole of the propagator (3.4) in his paper is exactly at the frequency of the upper-hybrid mode that we obtained above. The frequency of this mode in the $q\to 0$ limit is given in his equation (3.5) and reads 
$\omega_{\rm uh}\approx \omega_c +\frac{1}{2}[l_B^2n_0 v^{2D}(q)+\vf^2/2\omega_c]q^2$ in agreement with Eq.~(\ref{Eq:UH}).

\section{Discussion}

In this section we comment on the different regimes included in the theory. We also compare the analytical results obtained in Sec. \ref{Sec:Hydro} to the numerical solution of the quantum polarization function within the RPA. We further discuss the $B=0$ and $B\ne 0$ cases, and comment on the limitations of the hydrodynamic theory. 

 \subsection{Validity of collisionless hydrodynamics}
Hydrodynamics is usually valid in the low-frequency $\omega \tau \ll 1$ and long-wavelength $q v_F \tau \ll 1$ limits (known as the hydrodynamic limit), where $1/\tau$ is the collision rate between carriers. Here we study a clean, doped and degenerate graphene sample, which behaves as a two-dimensional Fermi liquid.\cite{KC12} The electron-electron collision rate is $1/\tau_{ee}\sim k_B^2T^2/\epsilon_F$ following Landau's famous phase space argument. At low $T$ and  close to the Fermi surface, $\tau_{ee}\to 0$ and therefore $\omega \tau_{ee} \gg 1$, which is the collisionless limit, where hydrodynamics is expected to fail. The standard approach is to use Landau's kinetic equation for a Fermi liquid.\cite{PL80} Here we follow a phenomenological approach pioneered by F. Bloch,\cite{B33}  which consists of using hydrodynamics heuristically also in the collisionless regime of a Fermi liquid. This can not be microscopically and quantitatively exact, but is usually qualitatively correct to describe collective modes such as the plasmon.\cite{F73} However, as shown by  I. Tokatly and O. Pankratov\cite{TP99} and others [see e. g. section 5.3.3 of Ref. \onlinecite{W04}], hydrodynamics-like equations can nevertheless be derived for a Fermi liquid in the collisionless regime. These are more complicated than Bloch's collisionless hydrodynamics, as they include a pressure tensor that in addition to the diagonal pressure also contains an off-diagonal component describing deformations of the shape of the Fermi surface (shear-like or viscosity-like term). Such terms account for the zero sound of a Fermi liquid, for example.\cite{PL80} Here, we rely on the simple collisionless hydrodynamics of Bloch, knowing its limitations and in particular that the sound velocity appearing in the plasmon dispersion and that in the upper hybrid mode, if there is a magnetic field, is underestimated by the neglect of Fermi surface deformations.

In collisionless hydrodynamics, the specificity of graphene compared to a usual 2DEG only enters in the replacements of the mass $m$ by the cyclotron mass $\ef/\vf^2$, of the equation of states $P=n \ef/2 \propto n^2$ by $P=n \ef/3 \propto n^{3/2}$, and of the Fermi energy $\ef=\kf^2/2m\propto n$ by $\ef=\vf \kf \propto \sqrt{n}$. These three modifications only depend on the dispersion relation in zero magnetic field. As we have seen in the previous section, one obtains nevertheless for both graphene and the 2DEG the same coefficient $\gamma=1/2$ describing the hydrostatic pressure of the electron gas in the collisionless limit.
In other words, only the linear spectrum of graphene is taken into account but not any chirality effects coming from the eigenvectors of the graphene Hamiltonian. This is an approximation, which is certainly wrong when approaching the neutrality point.

\begin{figure*}[htbp]
\begin{center}
\includegraphics[width=1.03\textwidth]{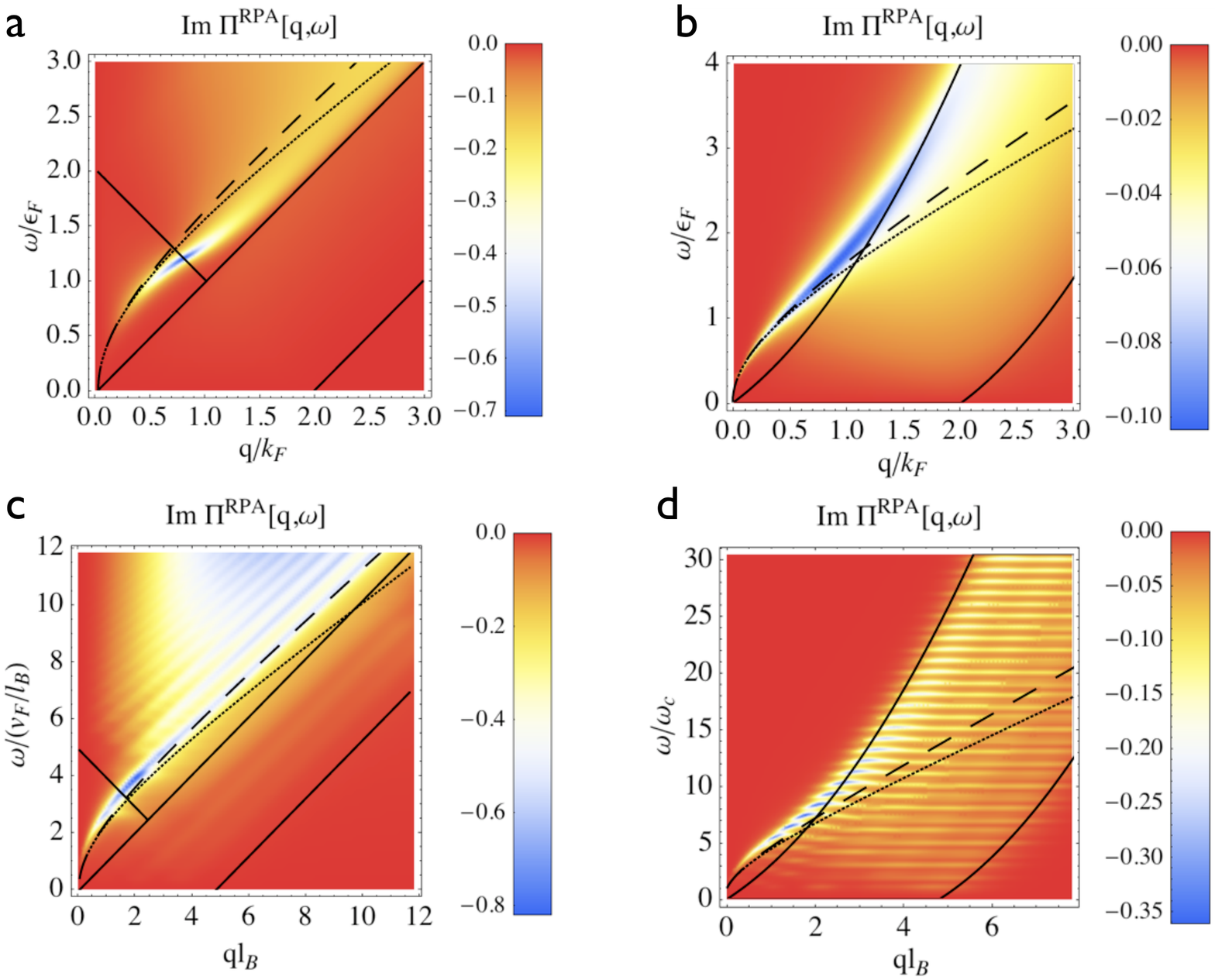}
\caption{(Color online) (a) Polarization spectrum of graphene within the RPA at $B=0$, and fitting of the plasmon mode to the dispersion relation Eq. (\ref{Eq:PlasmonB0}). (b) Same as (a) but for a 2DEG of massive electrons. (c) Same as (a) but in the presence of a strong magnetic field applied perpendicular to the graphene layer. The fitting of the upper hybrid mode is done using Eq. (\ref{Eq:UH}). (d) Same as (c) but for an standard 2DEG. In all the plots, the dotted line is the fitting to the hydrodynamical results Eq. (\ref{Eq:UH})-(\ref{Eq:PlasmonB0}) with $\gamma=1/2$, whereas the dashed line represents the fitting using the coefficient of the $q^2$ term $\gamma=3/4$, as obtained from expanding the 2DEG RPA analytical result up to $q^2$ order. See discussion in the text. The strength of the electron-electron interaction is plots (a) and (c) is $r_s= 1$, whereas $r_s= 3$ for plots (b) and (d). }
\label{Fig:Plasmons}
\end{center}
\end{figure*}

\subsection{Comparison with RPA}

A benchmark for the validity of the above classical hydrodynamic approach is the comparison with the plasmon obtained from the quantum RPA method. In general, the dispersion relation of collective plasmon modes can be calculated from the poles of the interacting polarization function.\cite{GV05} In Fig. \ref{Fig:Plasmons} we show density plots of the polarization function, $\Pi^{RPA}(\bq,\omega)$, within the RPA, \cite{RGF10} as well as the analytical solution of the upper hybrid mode dispersion relation Eq. \ref{Eq:UH} (dotted line) and the analytical approximation for the dispersion within RPA up to second order in $q$ (dashed lines).  Fig. \ref{Fig:Plasmons}(a) corresponds to the graphene spectrum at $B=0$. As expected, Eq. (\ref{Eq:UH}) properly reproduces the exact numerical RPA results at small wave vectors. Something similar is obtained for the $B\ne 0$ case of Fig. \ref{Fig:Plasmons}(c), for which the analytical approximation fit properly the poles of $\Pi^{RPA}(\bq,\omega)$ corresponding to the upper hybrid mode. However, the hydrodynamic solution, as given by the dotted black line, deviates from the numerical RPA solution at larger wave vectors. 

Indeed, the RPA does not only take into account the particular deformation of the Fermi surface due to density fluctuation, which correspond to the hyrdostatic pressure of the quantum gas, but also other non-local corrections. As such one may invoke the volume-preserving shear deformation of the Fermi surface mentioned above.\cite{GV05} This is a well known problem of the hydrodynamic approach and it is due to the local equilibrium assumption, which is inaccurate for plasmon modes.\cite{GV05} In fact, each plasmon oscillations involves a change in the shape of the Fermi surface, which costs additional kinetic energy.
Furthermore, one needs to take into account quantum corrections due to the wave-function overlap between the electron-hole pairs intervening in the plasmonic excitations. The latter gives rise namely to a particular chirality factor in the polarization function of graphene\cite{S86,WSSG06,HS07,RGF10} that reflects, among other effects, the absence of backscattering. A small-wavevector expansion beyond the classical term of the plasmon pole in the RPA polarization function yields a different result for the 2DEG, namely $\gamma=3/4$, that is larger than the result obtained from the hydrodynamic approach, as compared to graphene, where one finds a correction $\gamma=3/4-r_s^2$,\cite{PPV09} where $r_s=e^2/\epsilon_b \vf\simeq 2.2/\epsilon_b$ is the graphene fine-structure constant. However, in the case of graphene, this correction is hardly visible for physically relevant values of $r_s$ and becomes important only for $r_s\gg 1$, as we have checked (results not shown). In the present case of Fig. \ref{Fig:Plasmons} c, where we have plotted the imaginary part of the $B\neq 0$ RPA polarizability for a value of $r_s=1$, we have used simply $\gamma=3/4$ (for the dashed line) as a fitting parameter.

Apart from the upper hybrid mode, in the density plot of Fig. \ref{Fig:Plasmons}(c) we can observe an additional set of diagonal lines of strong spectral weight. These are the so called {\it linear magneto-plasmons}, and they have a fully quantum mechanical origin.\cite{RFG09,RGF10} The peculiar Landau level quantization of graphene into non-equidistant Landau levels permits an increasing number of inter-LL excitations and an enhancement of the LL mixing effect, due to the stacking of different Landau levels and the possibility of having transitions from the filled valence band. Those combined effects lead to a particular excitation spectrum of graphene as compared to that of the 2DEG with a parabolic band dispersion.\cite{RGF10} However, as we have mentioned, this is an intrinsic quantum-mechanical effect  due to the overlap between the wave functions of the intervening electrons and holes and therefore, it is not captured by the {\it semiclassical} approach of Sec. \ref{Sec:Hydro}.

We finish by highlighting some aspects of  the spectrum for a standard 2DEG of massive electrons. Fig. \ref{Fig:Plasmons}(b) shows the spectrum at $B=0$ from the exact RPA solution (density plot) and the analytical approximations within the hydrodynamic (dotted) and the RPA (dashed lines) approximations.\cite{RGF10}  In this case, the parabolic dispersion of massive single-band electrons leads to a different shape of the excitation spectrum. As before, the plasmon dispersion is well fitted, at small wave-vectors, by the analytical approximations. If a quantizing magnetic field is applied perpendicular to the 2DEG, then a set of equidistant LLs appears in the spectrum, separated by the cyclotron frequency $\omega_c=eB/m$, where $m$ is the  band mass, which coincides with the cyclotron mass (\ref{Eq:Cmass}) for a parabolic dispersion relation. In this case, the spectrum is discretized into \textit{weakly dispersing} horizontal lines placed at energies proportional to $\omega_c$, clearly seen in Fig. \ref{Fig:Plasmons}(d). As for the case of graphene, an upper hybrid mode appears also in the spectrum, which can be approximately described, in the long wavelength limit, by the analytical approximations discussed above.

\section{Conclusions}

In summary, we have used a simple collisionless hydrodynamic approach to study the small-wavevector collective modes of graphene and related materials, in the presence of a magnetic field perpendicular to the layer. The dispersion relation for the modes obtained this way has been compared to the quantum RPA result. We have further discussed the cases at zero field, as well as the more standard case of a 2DEG of massive electrons with a parabolic band dispersion. We show that, whereas the two approximations coincide at the leading order, the hydrodynamic approach fails to capture the correct behavior at $q^2$ order due to the local equilibrium assumption of this approximation.

\acknowledgements
R. R. acknowledges financial support from the Juan de la Cierva Program (MINECO, Spain). This work was partially funded by Triangle de la Physique.

\appendix

\section{Alternative derivation of the conservation laws for graphene}\label{AppendCons}
We start from Bloch's hydrodynamics equations
\begin{eqnarray}
\dot{\bf k}&=&-\frac{{\pmb \nabla} P(\br,t)}{n(\br,t)} -e{\bf E}-\frac{e}{c}{\bf v}(\br,t)\times{\bf B},\label{NewtonApp}\\
\frac{\partial n(\br,t)}{\partial t}&=&-{\pmb \nabla}\cdot[n(\br,t){\bf v}(\br,t)],\label{Eq:ContApp}
\end{eqnarray}
which can be expressed in terms of the Fermi velocity and the Fermi energy $\ef$ as 
\begin{equation}
{\bf v}_{\bk}={\pmb \nabla}_{\bk}\varepsilon_{\bk}=\vf\frac{\bk}{|\bk|}
\end{equation}

Taking into account that $\ef=\vf \kf$ we can express the wavevector as 
\begin{equation}
\bk=\frac{{\bf v}_{\bk}}{\vf} |\bk|\simeq \frac{{\bf v}_{\bk}}{\vf} \kf=m_c {\bf v}_{\bk}
\end{equation}
where $m_c\equiv \kf/v_F=\ef/v_F^2$ is the density dependent cyclotron mass of graphene. 
This allows us to write the first term of Eq. (\ref{Eq:Euler}) in the present form.

\bibliography{BibliogrGrafeno}

\end{document}